\documentclass[aps,pre,twocolumn,superscriptaddress]{revtex4-1}
\usepackage{graphicx,color,colortbl}
\usepackage{epstopdf}
\usepackage{bm}
\usepackage{eucal}
\usepackage{mathrsfs}

\usepackage[dvipsnames]{xcolor}
\usepackage{hyperref}
\definecolor{hypcolor}{named}{BlueViolet}
\hypersetup{colorlinks=true, citecolor=hypcolor, urlcolor=hypcolor, linkcolor=hypcolor}

\usepackage{amsmath}
\usepackage{amsthm}
\usepackage{epsfig}
\usepackage{lipsum}

\newcommand{\pdesreal}{729}
\newcommand{\pdesgauss}{819}
\newcommand{\ti}{\ensuremath{\theta_i}}
\newcommand{\tj}{\ensuremath{\theta_j}}
\newcommand{\avg}[1]{\ensuremath{\left<#1\right>}}
\newcommand{\tid}{\ensuremath{\dot{\theta}_i}}
\newcommand{\tidd}{\ensuremath{\ddot{\theta}_i}}
\newcommand{\omgg}{\ensuremath{\omega_\text{G}}}
\newcommand{\lmax}{\ensuremath{\lambda_1}}
\newcommand{\amat}{\ensuremath{A_{ij}}}

\newcommand{\Preal}{\ensuremath{P^{\mathrm{R}}}}
\newcommand{\Pgauss}{\ensuremath{P^{\mathrm{G}}}}

\newcommand{\dev}{\ensuremath{\Delta\omega}}
\newcommand{\deva}{\ensuremath{\avg{\dev}_t}}
\newcommand{\abs}[1]{\ensuremath{\left|#1\right|}}
\newcommand{\ncontrolled}{\ensuremath{N_\mathrm{c}}}

\begin{document}
	
	\title{Enhancing power grid synchronization and stability through time delayed feedback control }
	\author{Halgurd Taher}
	\affiliation{Inria Sophia Antipolis M\'{e}diterran\'{e}e Research Centre, 2004 Route des Lucioles, 06902 Valbonne, France}
	\affiliation{Institut f{\"u}r Theoretische Physik, Technische Universit\"at Berlin, Hardenbergstra\ss{}e 36, 10623 Berlin, Germany}
	\author{Simona Olmi}
	\email[corresponding author: ]{simona.olmi@fi.isc.cnr.it}
	\affiliation{Inria Sophia Antipolis M\'{e}diterran\'{e}e Research Centre, 2004 Route des Lucioles, 06902 Valbonne, France}
	\affiliation{CNR - Consiglio Nazionale delle Ricerche - Istituto dei Sistemi Complessi, 50019, Sesto Fiorentino, Italy}
	\author{Eckehard Sch{\"o}ll}
	\affiliation{Institut f{\"u}r Theoretische Physik, Technische Universit\"at Berlin, Hardenbergstra\ss{}e 36, 10623 Berlin, Germany}
	\date{\today}

	\begin{abstract}
We study the synchronization and stability of power grids within the Kuramoto phase oscillator model with inertia
with a bimodal frequency distribution representing the generators and the loads. The Kuramoto model
describes the dynamics of the ac voltage phase, and allows for a comprehensive understanding of fundamental network
properties capturing the essential dynamical features of a power grid on coarse scales. We identify critical nodes through solitary frequency 
deviations and Lyapunov vectors corresponding to unstable Lyapunov exponents. To cure
dangerous deviations from synchronization we propose time-delayed feedback control, which is an efficient control
concept in nonlinear dynamic systems. Different control strategies are tested and compared with respect to the
minimum number of controlled nodes required to achieve synchronization and Lyapunov stability. As a proof of
principle, this fast-acting control method is demonstrated for different networks (the German and the Italian power transmission grid), operating points, configurations, and models.
In particular an extended version of the Kuramoto model with inertia is considered, that includes the voltage dynamics, thus taking into account the interplay of amplitude and phase typical of the electrodynamical behavior of a machine. 

	\end{abstract}
	
\pacs{05.45.Xt, 87.18.Sn, 89.75.-k}
\keywords{nonlinear complex networks, power grids, synchronization, stability analysis, time-delayed feedback control}	
	
	\maketitle
	\section{Introduction}
	Synchronization phenomena in nonlinear dynamical networks are of major interest to a wide field of applications in natural and technological systems \cite{PIK01,BOC18}, 
	e.g., neural networks in the human brain, or supply and communication networks and power grids, which naturally have a strong link to economy. Research in these fields has revealed diverse phenomena 
	related to synchronization, ranging from partial synchronization patterns to asynchronous states \cite{van1996,VRE00,strogatz2001}. In particular, scenarios leading from full synchronization to asynchronicity 
	via \textit{solitary} states, i.e., single nodes which are desynchronized from the rest, play an important role for complex dynamical systems \cite{MAI14a,JAR18}, and in this work we will 
	show that they are fundamental also for power grids.
	
	Infrastructure, e.g., public transportation, medical care and a vast number of other everyday life applications, rely on electrical power supply. Given the fact that modern power transmission grids, 
	notably if they include renewable energy sources, differ significantly from conventional power grids with regards to topology and local dynamics \cite{MIL13,HEI10b,HEI11c}, it is necessary to identify, 
	understand, and cure the arising challenges and problems. In particular, malfunctioning grids can be the result of power outages, which occur for various reasons, including line overload or 
	voltage collapse. Here we will focus on the loss of synchrony. In normal operation, a power grid runs in the synchronous state in which all frequencies equal the nominal frequency (50 or 60 Hz) 
	and in which steady power flows balance supply and demand at all nodes. When parts of a power grid desynchronize, destructive power oscillations emerge. To avoid damage, affected components must 
	then be switched off. However, such switchings can in turn desynchronize other grid components, possibly provoking a cascade of further shut-downs and ending in a large-scale 
	blackout \cite{UCT1,MOT02,BUL10}. 
	
	The failure of a transmission line during a blackout can be determined not only by the network topology and the static distribution of electric flow but also by the collective transient dynamics of the entire system where the time scale of system instabilities is of seconds \cite{SCH18c,SIM08}. In general, grids are designed such that the synchronous state is locally stable, implying that a cascade-triggering desynchronization cannot be caused by a small perturbation. However, even if the synchronous state is stable against small perturbations, the state space of power grids is also populated by numerous stable non-synchronous states to which the grid might be driven by short circuits, fluctuations in renewable energy generation or other large perturbations  \cite{CHI10,ANV16,SCH18c}. Therefore it is of fundamental interest to explore the relation between network properties and grid stability against large perturbations  \cite{SCH15g, SCH18c, MEN13}. Yet many intriguing questions on the relation between grid topology and local stability are still not understood. Decentralized grids tend to be less robust with respect to dynamical perturbations, but more robust against structural perturbations to the grid topology \cite{ROH12}. However, adding new links may not only promote but also destroy synchrony, thus inducing power outages when geometric frustration occurs  \cite{WIT12,TCH18}. The local stability can be improved by relating the specifics of the dynamical units and the network structure  \cite{MOT13a, DOE13, MEN14}, or predicting \textit{a priori} which links are critical via the link's redundant capacity and a renormalized response theory \cite{WIT16}. 
	
    In this paper we will demonstrate the role played by the solitary nodes in driving the populations out of synchrony and the necessity to control these nodes when restoring both stability and synchronization.
    Solitary nodes can be related to local instabilities via the application of a standard stability toolbox (i.e., Lyapunov exponents and Lyapunov vectors), and to topological properties of the network, like dead ends, thus complementing the analysis reported in \cite{MEN14}. Once we have identified the critical power grid nodes which undermine stability and synchronization, we will apply time-delayed feedback control to a small subset of these nodes, in order to cure a desynchronized and unstable power grid. Time-delayed feedback is an efficient mechanism known in nonlinear dynamics and often used to control unstable systems \cite{PYR92,SCH07}.	Generator and consumer dynamics will be described in terms of: i) \textit{Kuramoto oscillators with inertia} \cite{FIL08a}; ii) extended model of Kuramoto rotators with non trivial voltage dynamics or \textit{synchronous machines} \cite{SCH14m}. As a specific example, we consider the topology of the German ultra-high voltage  power transmission grid (220~kV and 380~kV).
	
	\section{Model and methods}
	
	\subsection{The Kuramoto model with inertia}
	The Kuramoto model with inertia describes the phase and frequency dynamics of $N$ coupled synchronous machines, i.e., generators or consumers within the power grid, where mechanical and electrical phase and frequency are assumed to be identical: 
	
	\begin{align}
	\tidd+\alpha\tid=\frac{P_i}{I_i\omgg}+\frac{K}{I_i\omgg}\sum_{j=1}^{N}\amat\sin(\tj-\ti),\label{eq:kura}
	\end{align}
	
	with the phase $\ti(t)$ and frequency $\tid(t)=\frac{d\ti}{dt}$ of node $i=1,...,N$. Both dynamic variables $\ti(t)$, $\tid(t)$ are defined relative to a frame rotating with the reference power line frequency $\omgg$, e.g., 50~Hz for the European transmission grid. The distribution of net power generation ($P_i>0$) and consumption ($P_i<0$) is bimodal; it corresponds to the inherent frequency distribution in the Kuramoto model with rescaled parameters (see Appendix A for a detailed discussion on the parameter selection). 
	The power balance requires $\sum_i P_i =0$.We assume homogeneously distributed transmission capacities $K$. The adjacency matrix $\amat$ takes values 1 if node $i$ has a transmission line connected to node $j$, and 0 otherwise. Moreover $\alpha$ is the dissipation parameter and takes typical values of 0.1-1 s$^{-1}$ \cite{MEN14,MAC08a}. Finally, the moment of inertia $I_i$ of turbine $i$ is $I_i=I=40\times10^3 \text{kg}\text{ m}^2$, corresponding to generation capacities of a single power plant equal to 400~MW \cite{MEN14,HOR08}. With the above definitions, the frequency synchronization criterion reads $\tid(t)=0,\quad \forall i=$1,...N, i.e., deviations from the reference frequency are zero. 
	
	\subsection{Synchronous machine}
	Eq. (\ref{eq:kura}) has been derived in \cite{FIL08a} from the swing equation governing the rotor's mechanical dynamics \cite{MAC08a}, by assuming constant voltage amplitude and constant mechanical power $P_i$. The former assumptions make the model incapable of modeling voltage dynamics or the interplay of amplitude and phase. However it is possible to extend the model straightforwardly by including the voltage dynamics, thus taking into account the machine's electrodynamical behavior. In the following we consider a lossless network of synchronous machines, whose dynamics is described by the extended model derived in \cite{SCH14m}. The coupled dynamics of the phases $\left\lbrace \theta_i\right\rbrace $ and magnitudes $\left\lbrace E_i\right\rbrace$ of the complex nodal voltages $\left\lbrace \textbf{E}_i=E_i e^{i \theta_i}\right\rbrace _{i\in\{1,...,N\}}$ are given by
    
    \begin{align}
    \tidd+\alpha\tid&=\frac{P_i}{I_i\omgg}+\frac{K}{I_i\omgg}\sum_{j=1}^{N}\amat E_i E_j \sin(\tj-\ti),\label{eq:k}\\
    m_v\dot{E}_i&=-E_i+E_{f,i}+X_i\sum_{j=1}^{N}\amat   Ej\cos (\tj-\ti),\label{eq:v}
    \end{align}	

    where $\dot{\theta}_i$ is the individual frequency of the $i-$th oscillator. $P_i$ denotes the mechanical input or output power and $K A_{ij} E_i E_j \sin(\tj-\ti)$ is the electrical real power transferred between machines $i$ and $j$. The susceptance matrix coefficients $A_{ij}$ allow for variations concerning the network topology; as for the previous model, $A_{ij}=1$ if node $i$ has a transmission line connected to node $j$, 0 otherwise. In particular the diagonal entries of $\amat$ are chosen such that the matrix has zero row sum $\sum_{i=1}^N\amat=0$. 
$m_v$, $E_{f,i}$, $X_i$ take into account machine and line parameters. In particular these parameters are set to be homogeneous and of the same order of magnitude as in \cite{SCH14m}: 
\mbox{$m_v=1$}, \mbox{$E_{f,i}=1$}, \mbox{$X_i=1$}, while the remaining quantities, already discussed in the original model Eq.~(\ref{eq:kura}), are chosen as 
\mbox{$\alpha=5/6$~s$^{-1}$}, \mbox{$I_i=I=40\cdot 10^3 \text{kg}\text{ m}^2$}, \mbox{$\omgg=2\pi \cdot 50~$~Hz}.

	\subsection{German power grid and power distributions} 
	\label{sec.topology}
\begin{figure}[ht!]
		\includegraphics[width=0.9\linewidth]{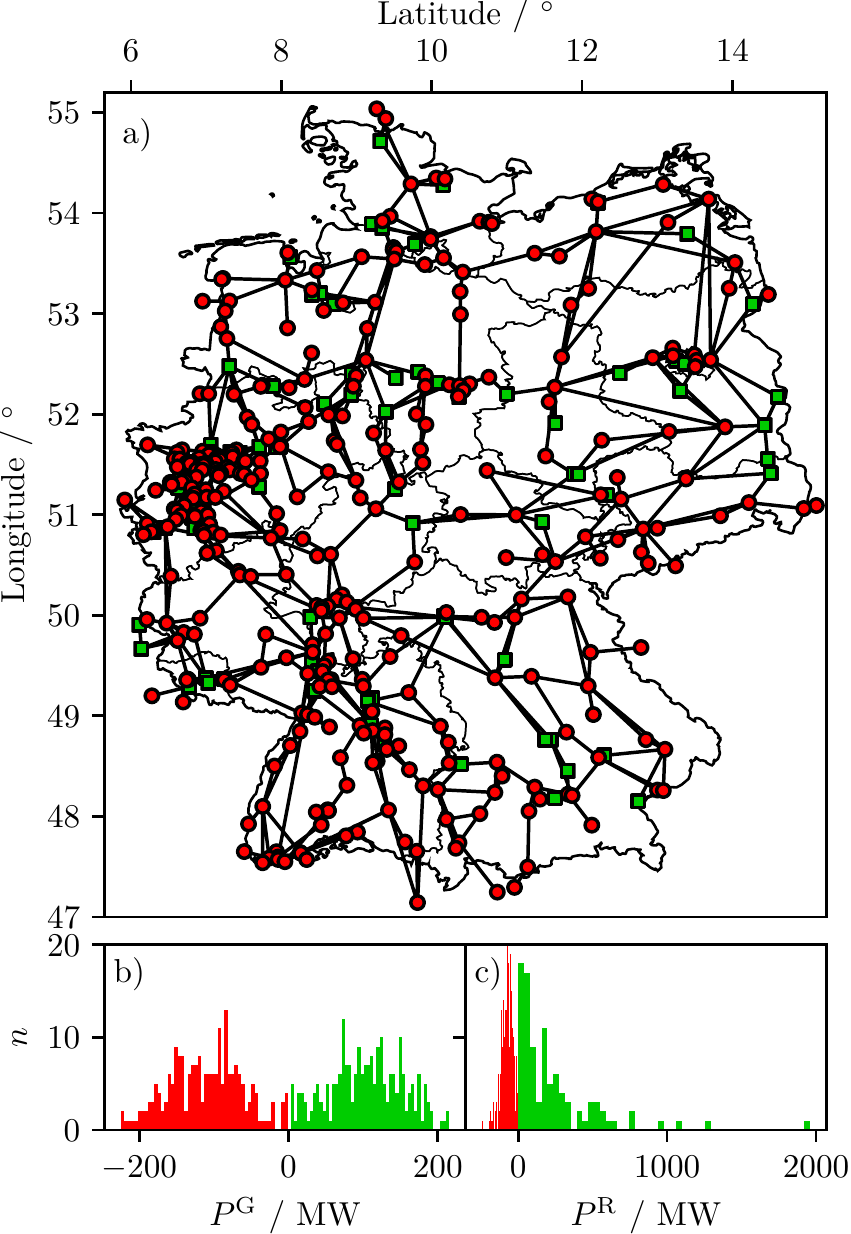}
		\caption{(a) Map of the German ultra-high voltage power grid, consisting of 95 net generators (green squares) and 343 net consumers (red dots) connected by 662 transmission lines (black lines). 
		(b,c) Histograms showing distributions of net power: (b) realization of an artificial bimodal Gaussian; $P_0=105$~MW, $\sigma=P_0/2$. 
		(c) based on the German power grid \cite{EGE16}. The green (red) bars correspond to generators (consumers).}
		\label{fig:1}
	\end{figure}

	In our numerical example we extract the topology $\amat$ from the \textit{Open Source Electricity Model for Germany} (elmod-de) \cite{EGE16}, 
	which describes the German ultra-high voltage transmission grid using $N=438$ nodes connected by 662 transmission lines (see Figure \ref{fig:1}a). 

	In many previous studies using the Kuramoto model with inertia to model power grid networks, the distribution of net power generation and consumption $P_i$ is set to be a bimodal $\delta$-distribution 
	\cite{ROH12,ROH14,WIT12,LOZ12,MEN14,OLM14a}. Here we consider more complex distributions: 
	first of all, an artificial bimodal Gaussian distribution $\Pgauss$ \cite{OLM16a,TUM18} is generated, whose probability density function $p(P)$ is given by the superposition of two Gaussians centered at $\pm P_0$ with standard deviation $\sigma$
	\begin{align}
	\label{p_gaussian}
	p(P)=\frac{1}{2\sigma \sqrt{2\pi}}\left(e^{-\frac{(P-P_0)^2}{2\sigma^2}}+e^{-\frac{(P+P_0)^2}{2\sigma^2}}\right).
	\end{align}
	Figure \ref{fig:1}b shows a histogram of the realization $\Pgauss$ used in the numerical simulations of this study. 
	The second distribution $\Preal$ shown in Figure \ref{fig:1}c is calculated based on data provided by elmod-de \cite{EGE16} and will be referred to as \textit{real-world} distribution. 
	
\begin{figure}[ht!]
	\includegraphics[width=0.9\linewidth]{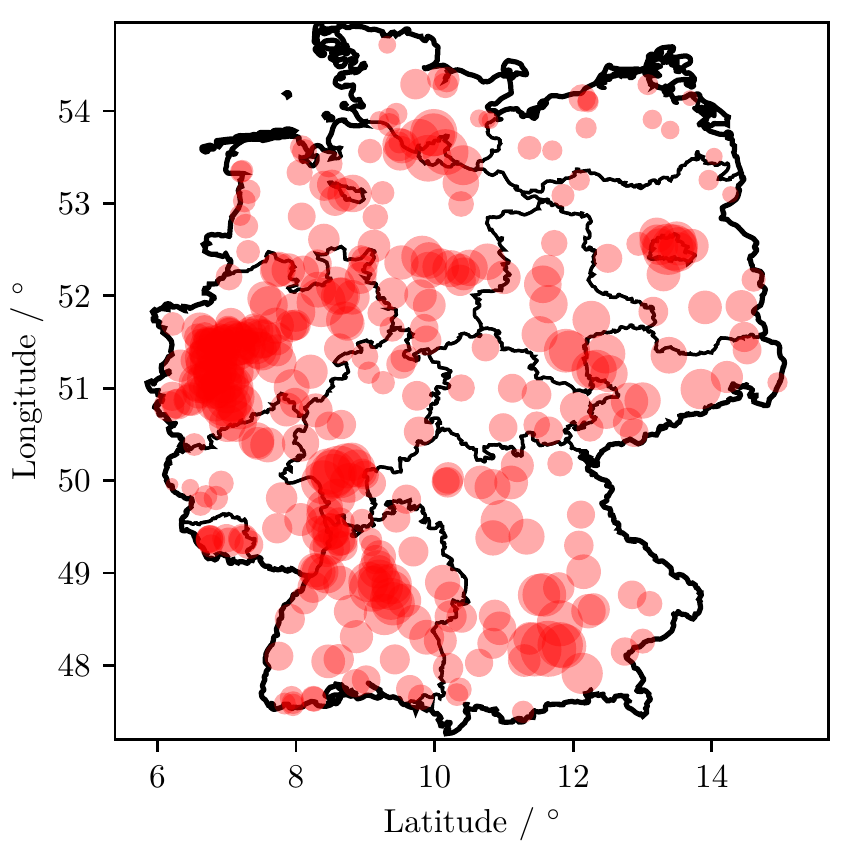}
	\caption{Spatial distribution of nodal power demands in elmod-de \cite{EGE16}. The size of the circles indicate the value of $p_i P_\text{Total}$.}
	\label{fig:s1}
\end{figure}

\begin{figure}[ht!]
	\includegraphics[width=0.9\linewidth]{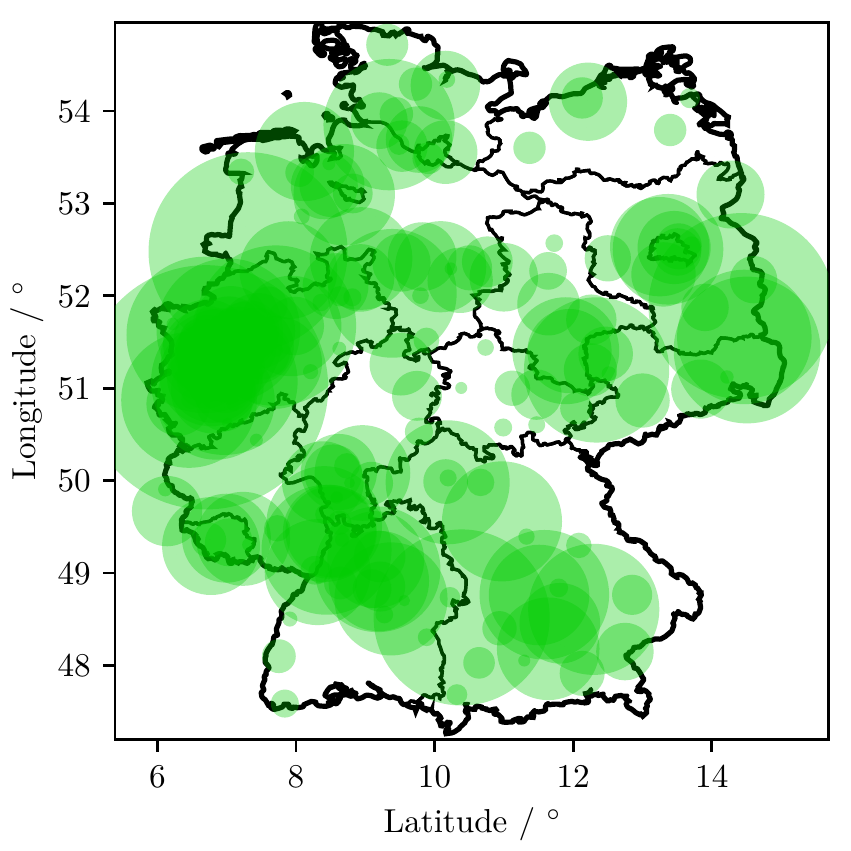}
	\caption{Spatial distribution of nodal generation capacities in elmod-de \cite{EGE16}. The size of the circles indicate the value of $C_i$.}
	\label{fig:s2}
\end{figure}

According to the data documentation \cite{EGE16}, elmod-de  is an open source nodal DC load flow model, minimizing generation costs, for the German electric power transmission grid. In the following, we point out how the information in elmod-de is translated into realistic values for the parameters used in our network of Kuramoto oscillators with inertia. 
As anticipated above, the data set contains nodal information on $N=438$ network nodes within the 220~kV and 380~kV ultra-high voltage transmission grid, of which 393 are substations. 
The remaining nodes are used to model interactions with neighboring countries (22) and auxiliary nodes (23), e.g., points in the grid without a transformer station. The nodes are connected with 697 transmission lines, 35 of them appearing twice in the data set, which will be neglected, such that 662 unique transmission lines remain. We will furthermore assume identical power transmission capacities for all transmission lines, resulting in a generic coupling strength for the network, thus reducing the values of the coupling matrix to 0 or 1. Besides geographical locations of all nodes, local power demand values $p_i$ are given in parts of the total power demand of Germany at off-peak times:
\begin{align}
P_\text{Total}=\sum_{i=1}^{N}p_iP_\text{Total}=36\text{~GW}
\end{align}
Following these definitions the absolute power demand at node $i=1,...,N=438$ is given by $p_i P_\text{Total}$. The spatial distribution of $p_i P_\text{Total}$ is illustrated in Fig. \ref{fig:s1}.
Furthermore 562 conventional power plants, e.g., coal or atomic plants, are listed. Information on the topological location of plants, i.e., to which node $i$ they belong,
and their maximum power generation capacities is provided. Let $n_i$ be the number of power plants associated with node $i$. The maximum capacity of plant $k=1,...,n_i$ located at node $i$ 
will be denoted by $C_i^k$. In order to obtain node-wise generation capacities $C_i$, $C_i^k$ will be aggregated for each node:
\begin{align}
C_i = 
\begin{cases}
\sum_{k=1}^{n_i}C_i^k	& \quad n_i>0\\
0 				 		& \quad n_i=0
\end{cases}
\end{align}
The spatial distribution of $C_i$ is illustrated in Fig. \ref{fig:s2}. The total generation capacity $C_\text{Total}$ reads:
\begin{align}
C_\text{Total}=\sum_{i=1}^N C_i=88.354~\text{GW}.
\end{align}
Due to the fact that plants being operated at 100\% of their maximum generation capacities would cause a large oversupply of power generation and break power balance, 
we will assume each plant to be operated at 41\% of its maximum capacity, since \mbox{$P_\text{Total}/C_\text{Total}\approx 0.41$}. With this intermediate level of power generation, 
the power balance is fulfilled and the net generation/consumption $P_i$ at node $i$ is given by:
\begin{align}
P_i=0.41\cdot C_i-p_i P_\text{Total}
\end{align}

	\subsection{Macroscopic indicators and Lyapunov analysis}
We consider a scenario where, due to an arbitrary dynamical perturbation, some critical nodes have become desynchronized, where we define as critical those nodes withstanding self-organized resynchronization. Synchronization is first gained by performing an adiabatic transition from the asynchronous to the synchronized state for increasing coupling constant: 
	starting with random initial conditions \mbox{$\ti(0)\in [-2\pi,2\pi)$}, \mbox{$\tid(0)\in[-1,1)$} at $K=0$, the coupling strength $K$ is increased adiabatically up 
	to $K_\mathrm{Max}$ where the system shows synchronized behavior. For each investigated value of $K$, the system is initialized with the final conditions
	found for the previous coupling value, then the system evolves for a transient time $T_A$, such that it can reach a steady state. 
	After the transient time $T_A$, characteristic measures are calculated in order to assess the quality of synchronization and the stability of the underlying state $\{\ti(T_A), \tid(T_A)\}$. 
	In particular the time-averaged phase velocity profile $\avg{\omega_i}_t\equiv\avg{\tid}_t$ provides information on frequency synchronization of individual nodes $i$, whereas the standard deviation of frequencies 
	\begin{equation}
	 \dev(t) \equiv \frac{1}{N}\sqrt{\sum_{i=1}^{N}\left(\omega_i(t)-\bar{\omega}(t)\right)^2},
	\end{equation}
	is used to estimate the deviation from complete frequency synchronization ($\bar{\omega}(t)$ indicates the instantaneous average grid frequency).

        Once a desired synchronized state is reached, a perturbation can occur leading the state out of synchrony. In this situation the overall stability of the power grid might be lost, therefore it is necessary to analyze the time-evolution of small dynamic perturbations $\delta\ti:=\ti^*-\ti$ around the steady state $\ti^*$, whose dynamics is ruled by the linearization of Eq. (\ref{eq:kura}) as follows
		\begin{align}
	\delta\tidd + \alpha\delta\tid = \frac{K}{I\omgg}\sum_{j=1}^{N}\amat \cos(\tj-\ti)(\delta\tj-\delta\ti).\label{eq:kuratangent}
	\end{align}	
	For the extended model, the linearization of Eqs. (\ref{eq:k}, \ref{eq:v}) reads as 
	\begin{eqnarray}
	\nonumber
	\delta\tidd + \alpha\delta\tid &=& \frac{K}{I\omgg}\sum_{j=1}^{N}\amat \left[ E_i E_j\cos(\tj-\ti)(\delta\tj-\delta\ti) \right.
	\\ \label{eq:kuraExttangent1}
	 &+& \left. (\delta E_i E_j +E_i \delta E_j) \sin(\tj-\ti) \right] 
	\\ \nonumber
	m_v\delta\dot{E}_i &=&-\delta E_i+X_i\sum_{j=1}^{N}\amat \left[ -E_j \sin(\tj-\ti) (\delta\tj-\delta\ti) \right. 
	\\ \label{eq:kuraExttangent2}
	 &+& \left.\delta Ej\cos (\tj-\ti)\right].
	\end{eqnarray}	
	The exponential growth rates of the infinitesimal perturbations are measured in term of the associated Lyapunov spectrum $\{\lambda_k\}$, with $k=1,...,2N$, numerically estimated by employing the method developed by Benettin et al. \cite{benettin1980}. 
        In particular one should consider for each Lyapunov exponent $\lambda_k$ the corresponding 2N-dimensional tangent vector $\mathcal{T}^{(k)}=(\delta\dot{\theta}_1,...,\delta\dot{\theta}_N, 
        \delta\theta_1,...\delta\theta_N)$ whose time evolution is given by Eq. (\ref{eq:kuratangent}) (resp. Eqs. (\ref{eq:kuraExttangent1}, \ref{eq:kuraExttangent2}) for the extended model). 
        Important information about the sources of instability and, in particular, about the oscillators that are more actively contributing to the chaotic dynamics, can be gained
        by calculating the time averaged evolution of the tangent vector $\mathcal{T}^{(1)}$, here referred to as maximum Lyapunov vector. The Euclidean norm of each $\{\ti,\tid\}$ pair in  $\mathcal{T}^{(1)}$, averaged in time, is measured for each oscillator as $\xi_i:= \left<\sqrt{[\delta\ti(t)]^2+[\delta\tid(t)]^2}\right>_t$,	once the tangent vector is orthonormalized, i.e. $||\mathcal{T}^{(1)}||=1$.


	\section{Results for a network of Kuramoto oscillators with inertia}
	
\subsection{Emergence of solitary states}
\begin{figure}[h]
	\centering
	\includegraphics[width=1\linewidth]{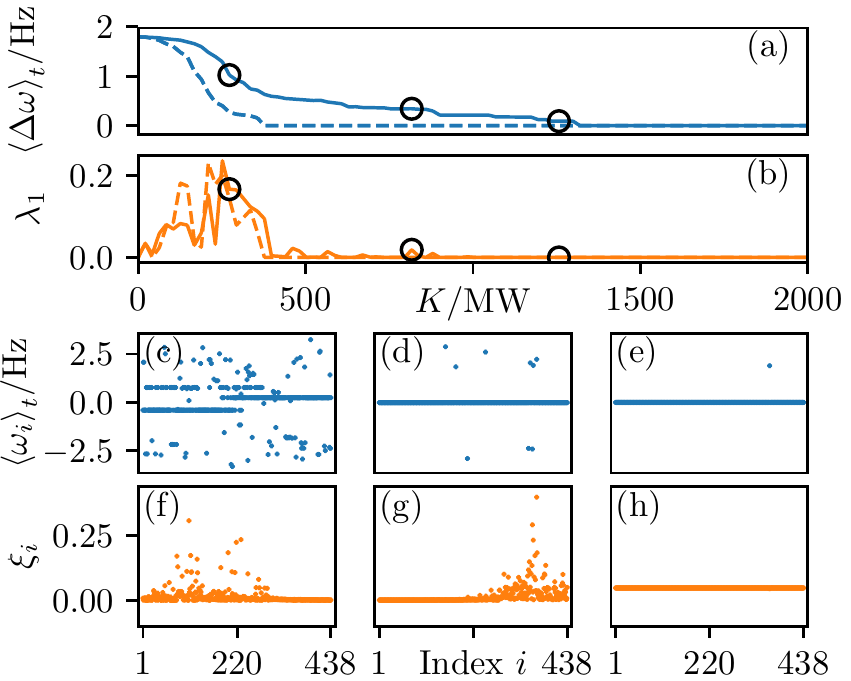}
	\caption{German power grid with bimodal Gaussian distribution $P^\mathrm{G}$: (a) Average frequency deviation $\avg{\Delta \omega}_t$ and (b) 
		largest Lyapunov exponent $\lmax$ versus coupling strength $K$. The solid (dashed) lines correspond to the adiabatic upsweep (downsweep) of $K$. 
		\mbox{(c) - (e)} Average frequencies $\avg{\omega_i}_t$ and \mbox{(f) - (h)} Lyapunov vector components $\xi_i$ versus node index $i$ for the $K$ values marked by black circles 
		in the top panels from left to right. Parameters: \mbox{$0 \le K \le 3142$~MW} in steps of \mbox{$\Delta K\approx21$~MW} with \mbox{$\alpha=5/6$~s$^{-1}$}, 
		\mbox{$I_i=I=40\cdot 10^3 \text{kg}\text{ m}^2$}, \mbox{$\omgg=2\pi \cdot 50~$~Hz}. Averages taken over 100~s after discarding a transient time of 400~s. Lyapunov exponents and vectors 
		calculated for a duration of $4\cdot 10^5$~s. Lyapunov exponents are expressed in units of $\Delta t^{-1}=5 s^{-1}$.}
	\label{fig:adynamicswoutctrlgergauss}
\end{figure}

\begin{figure}[h]
	\centering
	\includegraphics[width=1\linewidth]{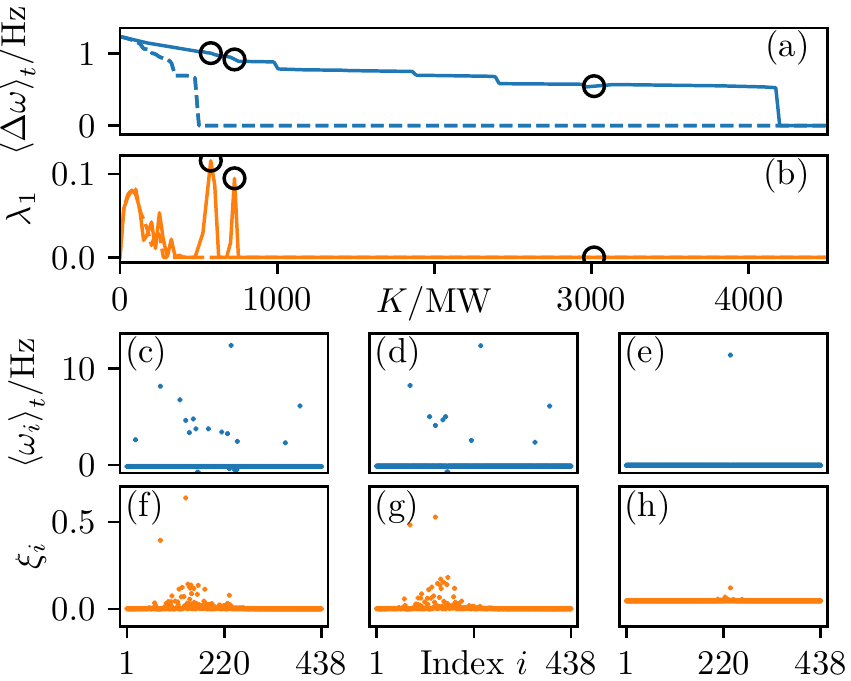}
	\caption{German power grid with real-world distribution $P^\mathrm{R}$: Same as in Fig. \ref{fig:adynamicswoutctrlgergauss}. Parameters: 
		\mbox{$0 \le K \le 4500$~MW} in steps of \mbox{$\Delta K\approx25$~MW} with \mbox{$\alpha=2$~s$^{-1}$}. Other parameters as in Fig. \ref{fig:adynamicswoutctrlgergauss}.}
	\label{fig:adynamicswoutctrlgerrealworld}
\end{figure}

In general we have performed sequences of simulations by varying adiabatically the coupling parameter $K$ with two different protocols. Namely, for the upsweep protocol, as described 
in the previous section, the series of simulations is initialized for the decoupled system by considering random initial conditions both for phases and frequencies.
Afterwards the coupling is increased in steps of $\Delta K$ until a maximum coupling strength $K_\mathrm{Max}$ is reached. 
For the downsweep protocol, starting from the maximum coupling strength $K_\mathrm{Max}$ achieved by employing the upsweep protocol
simulation, the coupling is reduced in steps of $\Delta K$ until $K = 0$ is recovered. At each step the system is simulated for a transient time $T_A$ followed by a time interval $T_W$ during which the average frequencies $ \left\langle \omega_i\right\rangle _t$, as well as the components of the Lyapunov vector $\xi_i$ and the maximum Lyapunov exponent $\lambda_1$ are calculated.
An example of the results obtained by performing the sequence of simulations of upsweep followed by downsweep is shown in Figs. \ref{fig:adynamicswoutctrlgergauss}  and \ref{fig:adynamicswoutctrlgerrealworld} for the bimodal Gaussian distribution $\Pgauss$ and the real-world distribution $\Preal$, respectively.

In both cases, at low coupling, a large fraction of the network is unsynchronized (panel (a)) and the system is chaotic, i.e., $\lambda_1>0$  (b). A considerable part of the oscillators
rotates with average frequency $\abs{\avg{\omega_i}}>0$, while relatively few oscillators are locked at average zero frequency (c). Other clusters at $\abs{\avg{\omega_i}}\neq0$ may emerge.
The \textit{solitary nodes}, which are desynchronized from the rest of the network, and oscillate with high frequency, are those mostly responsible for the lack of synchronization. 
This is revealed by the analysis of the components of the maximum Lyapunov vector $\xi_i$, which assume large values for those nodes which are solitary, thus indicating that the directions identified by solitary nodes are the most unstable in the network (as shown in panel (f)).

For intermediate $K$ values, the majority of nodes is synchronized on average, with a small set of nodes being solitary, for instance, 9 for the Gaussian and 11 for the real-world distribution
(see panels (a) and (d)). The system is still chaotic (panel (b)) and the components of the Lyapunov vector $\xi_i$ are still localized around solitary nodes (panel (g)).
The number of solitary nodes diminishes for increasing coupling values, since more and more nodes join the main synchronized cluster at zero average frequency.
Just before full synchronization (see panel (e)), one solitary node is left and no instability emerges in the system (panel (h)). The full synchronized state ($\left\langle \Delta\omega\right\rangle _t=0$) is stable and it is characterized by a single cluster with no solitary nodes. In particular complete frequency synchronization with $\deva=0, \lmax=0$ is achieved at $K\geq1320$~MW ($K\geq4200$~MW) for $\Pgauss$ ($\Preal$). 

When $K$ is decreased starting from the synchronized states, the systems remains synchronized for a larger $K$ interval, due to the hysteretic nature of the transition,
and the synchronized state loses stability (i.e., $\left\langle \Delta\omega\right\rangle _t>0$) for a coupling value smaller than the one found during the upsweep protocol (see panel (a)). 
The system is multistable and partially synchronized states (as those shown in panel (d)) coexist with the synchronized one.  
Depending on the initial state of the system, the dynamics can approach either the synchronized state or one of the upper branch states. This also means, that, starting from 
the synchronized states, large perturbations can kick the system out of synchrony. The goal of this paper it to give a proof of principle that once such a partially synchronized state is approached, our control method is capable of synchronizing and stabilizing the system. Thus in the following we consider the unstable states present in panels (d), (g) of Figures \ref{fig:adynamicswoutctrlgergauss} and \ref{fig:adynamicswoutctrlgerrealworld}, which we aim to control.

\subsection{Application of time-delayed feedback control}
\begin{figure}[ht!]
		\includegraphics[width=1\linewidth]{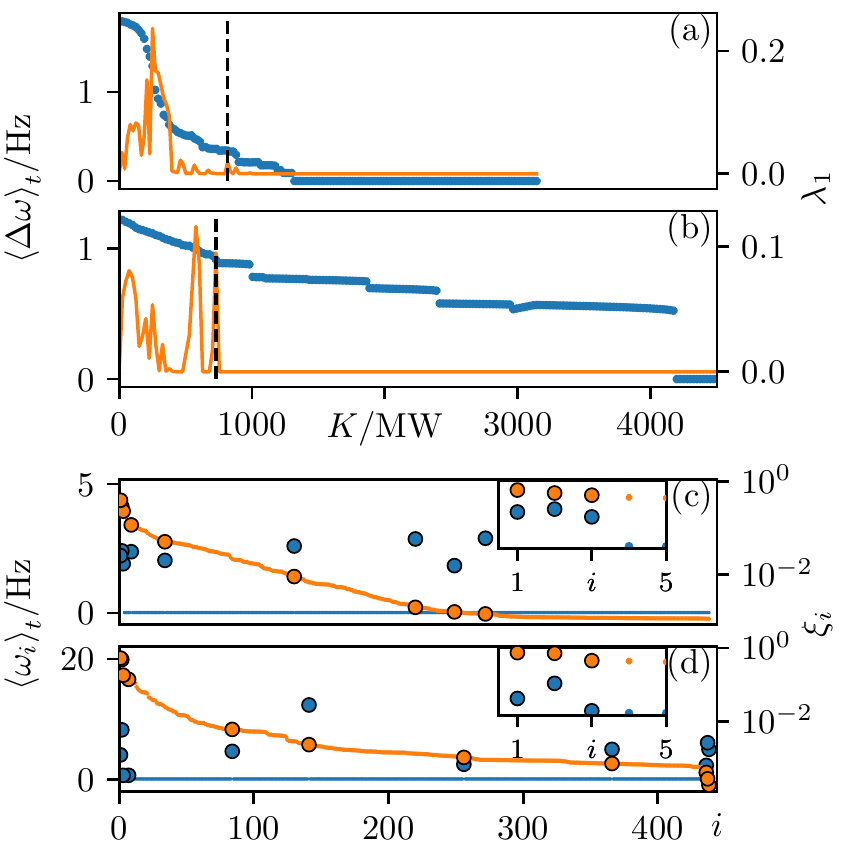}
		\caption{Time averaged standard frequency deviation $\deva$ (blue dots) and maximum Lyapunov exponent $\lmax$ (orange line) versus coupling strength $K$ for the bimodal Gaussian $\Pgauss$
		(a) and for the real-world distribution $\Preal$ (b), calculated for the upsweep protocol. The dashed black lines highlight the operating $K$ points. 
		Time averaged phase velocity profile $\avg{\omega_i}_t$ (blue dots) and Lyapunov vector components $\xi_i$ (orange dots) versus node index $i$ 
		for the bimodal Gaussian at \mbox{$K\approx\pdesgauss$~MW} (c) and for the real-world distribution at \mbox{$K\approx\pdesreal$~MW} (d). Data are ordered in descending order of $\xi_i$. 
		The insets show a zoom for small $i$. Large filled circles mark solitary nodes. For \mbox{$\Pgauss$} \mbox{$0 \le K \le 3142$~MW} in steps of \mbox{$\Delta K\approx21$~MW} 
		with \mbox{$\alpha=5/6$~s$^{-1}$}.  For \mbox{$\Preal$} \mbox{$0 \le K \le 4500$~MW} in steps of \mbox{$\Delta K\approx25$~MW} with \mbox{$\alpha=2$~s$^{-1}$} \cite{parameter}. 	
		Other parameters as in Fig. \ref{fig:adynamicswoutctrlgergauss}.
		}
		\label{fig:2}
	\end{figure}

To facilitate understanding we report in Fig. \ref{fig:2} the main features of the unstable states, briefly introduced in the previous section, that we aim to control.
In particular Fig. \ref{fig:2} shows the time-averaged standard frequency deviation $\deva$ and the maximum Lyapunov exponent $\lmax$ for each  value $K$ of the adiabatic increase for the bimodal Gaussian (panel a) and for the real-world distribution (panel b) and highlights the considered operating points via dashed black lines.

If a perturbation pushes the system out of synchrony at an intermediate state with finite values of $\deva$, in a chaotic regime characterized by $\lmax >0$, would it be possible to enhance synchronization and stability by controlling a small subset of nodes? In the following we will give a positive answer to this question, by exploring the dynamics of the system at $K\approx816$~MW ($K\approx729$~MW) for $\Pgauss$ ($\Preal$), where deterministic chaos is present, i.e., $\lmax=0.0187$ ($\lmax=0.096$), and the system is not perfectly frequency synchronized: $\mbox{$\deva\approx 0.34$~Hz}$ (\mbox{$\deva\approx0.91$~Hz}), modeling a strongly perturbed power grid \cite{pert}.
Even though we are considering a partially synchronized regime with an intermediate transmission capacity value, as a resulting regime in case of strongly perturbed grid, we made sure not to artificially drive the system to an unrealistic range of capacity values. Indeed the operating point at which we are working is in a realistic regime when considering the average transmission capacity ($\approx1500$ MW) at which the German ultra-high voltage transmission grid works, according to elmod-de data set. 
	
	From the average frequency profile shown in Fig. \ref{fig:2}, panel c (panel d) for $\Pgauss$ ($\Preal$), we can see that a major part of the power grid is frequency synchronized while few nodes have a significant frequency deviation and are identified as \textit{solitary states}: 9 nodes for $\Pgauss$, 11 nodes for $\Preal$. (Note that the three solitary nodes $i=1,2,3$ 
	can only be resolved in the blown-up inset.) Solitary nodes oscillate with their own average frequency and do not resynchronize in a self-organized way at a given coupling strength, 
	being thus critical for desynchronization. Note that the solitary nodes include those with the largest $\xi_i$, but not only those.

In order to enhance frequency synchronization and stability at the intermediate coupling strength discussed above, the Kuramoto model with inertia is now extended by time-delayed feedback control
which is an efficient control concept, well known in nonlinear dynamic systems \cite{PYR92,SCH07}, but also commonly employed in power grid engineering \cite{MAC08a, KUN94}:
	\begin{align}
	\tidd+\alpha\tid	&=\frac{P_i}{I\omgg}+\frac{K}{I\omgg}\sum_{j=1}^{N}\amat\sin(\tj-\ti)\nonumber\\
	&-\frac{g_i\alpha}{\tau}\left[\ti(t)-\ti(t-\tau)\right],
	\end{align}
	where $g_i$ is the control gain of node $i$ and $\tau$ is the delay time. The control method turns out to be robust against changes in the parameters $\tau, g_i$.
	We propose in the following to apply the control term only to a small subset of nodes selected according to their dynamical properties. 
	
\begin{figure}[ht!]
		\includegraphics[width=1\linewidth]{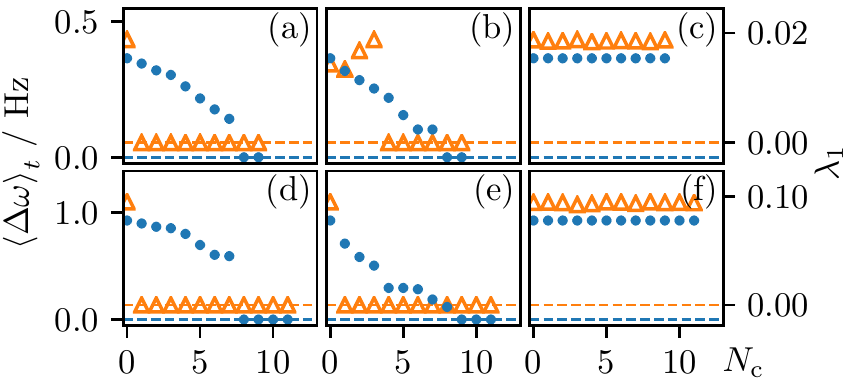}
		\caption
		{Efficiency of time-delayed feedback control: time averaged frequency deviation $\deva$ (blue dots) and maximum Lyapunov exponent $\lmax$ (orange triangles) vs. 
		number of controlled nodes $\ncontrolled$ following different control strategies: (a), (d) solitary nodes sorted in descending order of $\xi_i$. (b), (e) 
		solitary nodes sorted in descending order of $\abs{\avg{\omega_i}_t}$. (c), (f) randomly picked nodes. At each step of each control strategy, one more node
		is controlled, picked from one of the three mentioned lists, and both the level of synchronization and the instability are recalculated via $\deva$ and $\lmax$.
		Panels (a)-(c) correspond to the distribution $\Pgauss$, (d)-(f) to $\Preal$. 
		The dashed lines mark $\deva=0$, $\lmax=0$. Control acts for a duration of 40 seconds and is then turned off; delay time \mbox{$\tau=4$~s}, feedback gain $g=1$, 
		other parameters as in Fig. \ref{fig:adynamicswoutctrlgergauss}, time averages over 80 s.}
		\label{fig:4}
	\end{figure}

In order to find such a set, different control strategies are proposed in the following: (i) the first strategy takes into consideration all solitary nodes, sorted in descending order of $\xi_i$;  
(ii) the second strategy orders the solitary nodes by their absolute average frequency $\abs{\avg{\omega}_t}$; (iii) the third strategy consider all nodes, not only solitary ones, randomly picked.
	The outcome of the different strategies is shown in Fig.~\ref{fig:4}(a)-(c) and (d)-(f) for the bimodal Gaussian distribution $\Pgauss$ and the real-world distribution $\Preal$, respectively.
	First of all, strategy (i) is able to achieve stability if just one node is controlled, and frequency synchronization if the number of controlled solitary nodes is sufficiently large: 8 controlled nodes for both $\Pgauss$ and $\Preal$. 
Strategy (ii) requires 4 controlled nodes for stabilization and 8 for synchronization in case of $\Pgauss$, and one controlled node for stabilization and 9 nodes for synchronization in case of $\Preal$.
	The third strategy is not able to frequency-synchronize and stabilize, it can at most mitigate to some extent the desynchronization and the instability. 
	For the given setup, strategy (i) is the best choice: it is particularly efficient since the Lyapunov vector is re-calculated every time when an additional solitary node is controlled, 
	thus taking into account the interplay between solitary states and emerging instabilities. However, both strategies (i) and (ii) highlight the role played by solitary nodes, role that will be clarified in more details in the next section.
\subsection{Lyapunov analysis}	

The presence of solitary nodes deeply influences the dynamics emerging in the system, since they behave almost independently, adding complexity and conveying the instability.
In particular the role played by the solitary nodes can be understood by the change in the Lyapunov spectrum when the control strategy (i) is applied, i.e., when solitary nodes
are controlled, ordered according to their Lyapunov vector component (for the definition of the other strategies see previous section).

\begin{figure}[ht!]
	\includegraphics[width=1\linewidth]{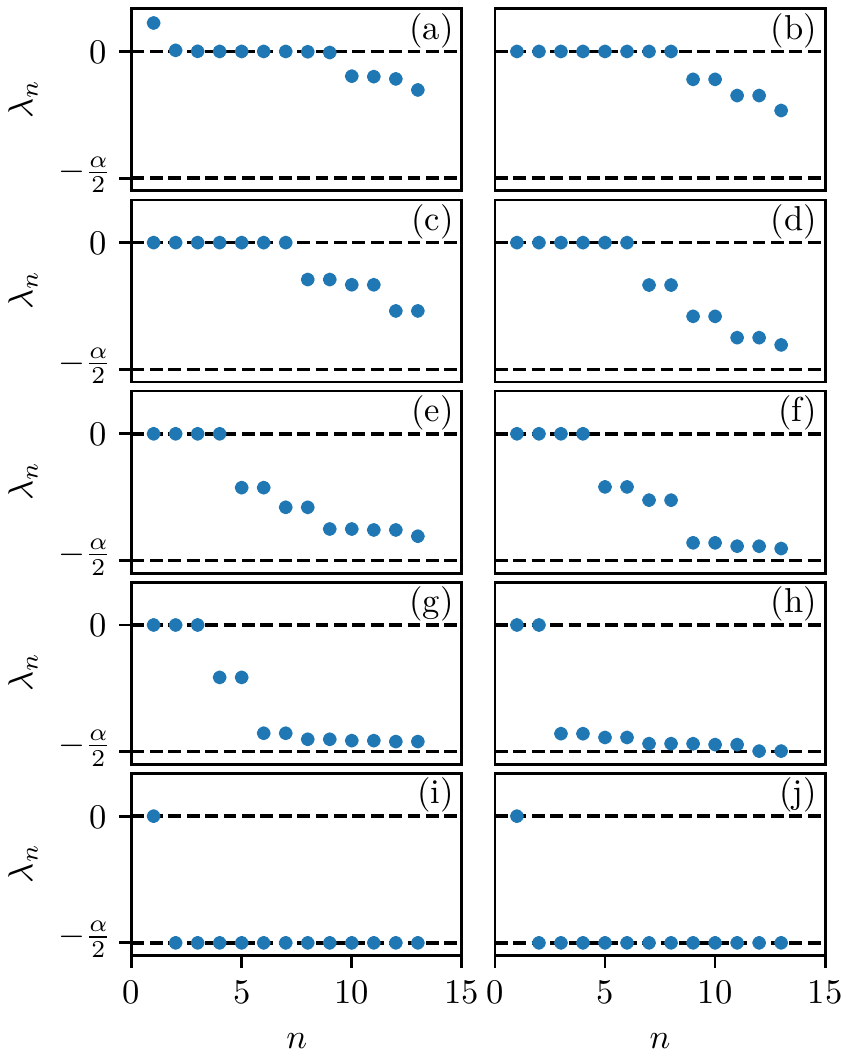}
	\caption{Bimodal Gaussian frequency distribution, control strategy (i): Lyapunov exponents $\lambda_n$ versus n for \mbox{$K= 819$ MW}. For simplicity only the first 13 exponents 
		of the spectrum are plotted. Panels (a) to (j) are arranged according to the number of controlled nodes $N_c$ increasing by one from $N_c = 0$ to $N_c = 9$.
		Lyapunov exponents are expressed in units of $\Delta t^{-1}=5 s^{-1}$.}
	\label{fig:s3}
\end{figure}

If we first consider the bimodal Gaussian frequency distribution, the uncontrolled state is characterized by a cluster of synchronized oscillators plus 9 solitary nodes. The system is chaotic
and the maximum Lyapunov exponent is positive (see Fig.~\ref{fig:s3}a): the interplay between solitary nodes and cluster state gives rise to low-dimensional chaos in the system.
When the first solitary node is controlled (Fig.~\ref{fig:s3}b), the dynamics becomes quasiperiodic and the collective behavior is a high-dimensional torus, as can be deduced by the 
consistent number of 8 Lyapunov exponents that are exactly zero. Each solitary node, at the microscopic level, moves with an average velocity which is different from the velocity of 
the cluster and from the velocity of the other solitary states: the self-emergent dynamics, at the macroscopic level, is a quasiperiodic motion characterized by multiple incommensurable 
frequencies. When solitary nodes are controlled and frequency synchronized to the cluster, they do no longer contribute to the collective dynamics with their own frequency, thus decreasing 
the dimensionality of the macroscopic behavior. Thus, the further control of more solitary nodes has the effect of stabilizing the system: negative exponents becomes more and more negative 
while the zero ones become negative. When 5 solitary states are controlled, the macroscopic dynamics evolves on a 2-dimensional torus (see Fig.~\ref{fig:s3}f). This can be explained 
considering that in the system under investigation one might expect two Lyapunov exponents to be zero due to the symmetries of the system: one is always present for a system with continuous 
time, while the second zero exponent is related to the invariance of the model under uniform phase shift. Therefore when 5 solitary nodes are controlled, 2 exponents are zero due to symmetries,
while the other 2 zero exponents identify the emergent quasiperiodicity.	
Finally, when the system is synchronized, thanks to the control of 8 solitary nodes, the typical spectrum of a stable periodic synchronized state appears, with a negative plateau at 
$\lambda_n=-\alpha/2$ (for $1<n<2N-1$) and $\lambda_1=0$ (see Fig.~\ref{fig:s3}i, j). The synchronized state is degenerate and the phase shift of all the phases corresponds to a 
perturbation along the orbit of the fully synchronized state, which explains why the two invariances, and thus the Lyapunov exponents, coincide, as already shown in \cite{OLM15b} 
for a globally coupled network.

\begin{figure}[ht!]
	\includegraphics[width=1\linewidth]{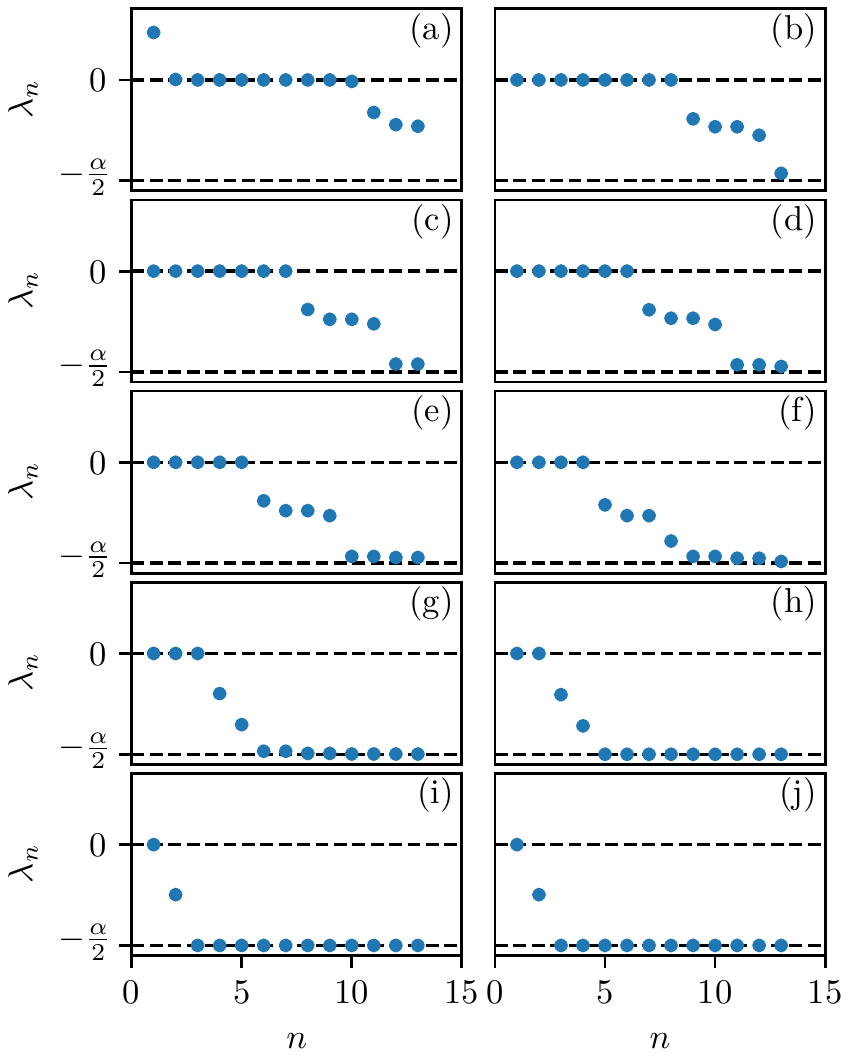}
	\caption{Real-world frequency distribution, control strategy (i): Lyapunov exponents $\lambda_n$ versus n for \mbox{$K= 729$ MW}. For simplicity only the first 13 exponents of the 
		spectrum are plotted. Panels (a) to (l) are arranged according to the number of controlled nodes $N_c$ increasing by one from $N_c = 0$ to $N_c = 11$.
		Lyapunov exponents are expressed in units of $\Delta t^{-1}=5 s^{-1}$.}
	\label{fig:s4}
\end{figure}

A similar behavior can be observed for the real-world frequency distribution case, where the initial uncontrolled state is chaotic ($\lambda_1>0$) and 11 solitary nodes emerge from the synchronized cluster state (see Fig. \ref{fig:s4} a). When the solitary state with largest Lyapunov component is controlled and synchronized to the cluster, the system is no longer unstable, which indicates that the instability was conveyed by the selected solitary node (see Fig.~\ref{fig:s4}b). Due to the interaction of the remaining solitary states, characterized by different
average frequencies, the collective dynamics of the system turns out to be quasiperiodic and high-dimensional. The dimensionality of the quasiperiodic motion is reduced by controlling more and more nodes and results in a 2-dimensional torus when 5 solitary nodes are controlled (see Fig.~\ref{fig:s4}f). Finally the system is synchronized when 8 solitary states are controlled (see Fig.~\ref{fig:s4}i),
while the additional control of further nodes does not alter nor enhance the synchronization.  

\subsection{Topological features vs Extreme events}	
\begin{figure}[ht!]
		\includegraphics[width=1\linewidth]{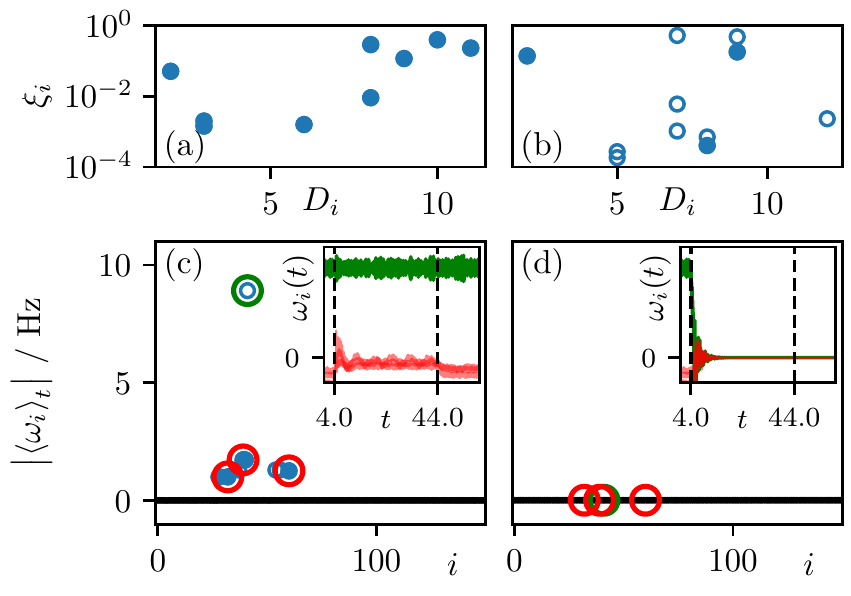}
		\caption
		{Source of solitary nodes: Lyapunov vector components $\xi_i$ versus maximum neighborhood degree $D_i$ for (a) $\Pgauss$, (b) $\Preal$. Only solitary nodes are shown, and filled circles identify nodes which belong to dead trees. (c),(d): Absolute time-averaged frequency $|\avg{\omega_i}_t|$ versus node index $i$ for $\Pgauss$, where $4\Delta P$ is added to the inherent frequency of an arbitrary non-solitary node $k$ (green circle). In (c) dead-tree nodes (red circles) adjacent to $k$ are controlled and in (d) $k$ is controlled. Black dots are synchronized nodes, blue symbols are solitary nodes. Nodes belonging to a dead tree are marked by filled symbols. The instantaneous frequencies $\omega_i(t)$ of green and red nodes versus time are shown in the insets. Vertical dashed lines mark activation and deactivation of control. Parameters as in Figs.4 and 7, time averages over 80 s.}
		\label{fig:topology}
	\end{figure}

	In \cite{MEN14} numerical evidence was given that dead ends and dead trees undermine basin stability of nodes in Kuramoto power grid networks, which means that the basin of 
attraction of the frequency synchronized solution for single nodes tends to be small if a node is placed at a dead end, thus making such nodes hard to synchronize. 
Indeed, in the case of the bimodal Gaussian distribution $\Pgauss$, all the identified solitary nodes belong to a dead tree (see Fig.~\ref{fig:topology}a). 
However, this trend cannot be observed for the real-world distribution $\Preal$, where just 3 of the 11 solitary nodes belong to a dead tree (see Fig.~\ref{fig:topology}b) and dead trees do not correspond
to the most unstable nodes. In general we have observed that the most unstable solitary nodes, for  $\Pgauss$, are dead ends adjacent to well connected nodes, whereas for $\Preal$ they are nodes with $P_i> 4\Delta P$, where $\Delta P$ is the standard deviation of the distribution. The discrepancy between the two cases can be explained if, starting from $\Pgauss$, we arbitrarily add $4\Delta P$ to the net power ($\widehat{=}$ inherent frequency) of a non-solitary node $k$. This altered node then becomes solitary and causes other adjacent nodes to become solitary, some of them belonging to dead trees. If we control all the newly emerged solitary dead trees, the system does not synchronize and the dynamics of node $k$ is almost unchanged (Fig. \ref{fig:topology}c), whereas we can achieve synchronization via controlling node $k$ only (Fig. \ref{fig:topology}d). This means that dead trees are fundamental in determining the power grid stability whenever the power distribution does not contain fat tails or \textit{extreme events}, which is the case for $\Pgauss$; for the real-world distribution $\Preal$, however, nodes with significant power difference are common and the stability is undermined by these nodes rather than by dead trees.

\section{Results for a lossless network of synchronous machines}
	
Applying the same procedure as previously done for the standard Kuramoto model with inertia with different frequency distributions, we perform an adiabatic parameter scan in $K$, thus identifying the synchronization transition of the system during the upsweep and downsweep protocols. The system is initialized at $K=0$ with uniformely distributed initial 
conditions not only for phases and frequencies $\{\theta_i, \dot{\theta}_i\}$, but also for the voltage amplitudes $\{E_i\}$, that are set uniformly random: $E_i(0)\in [0.5,1.5)$.
\begin{figure}[h]
	\centering
	\includegraphics[width=1\linewidth]{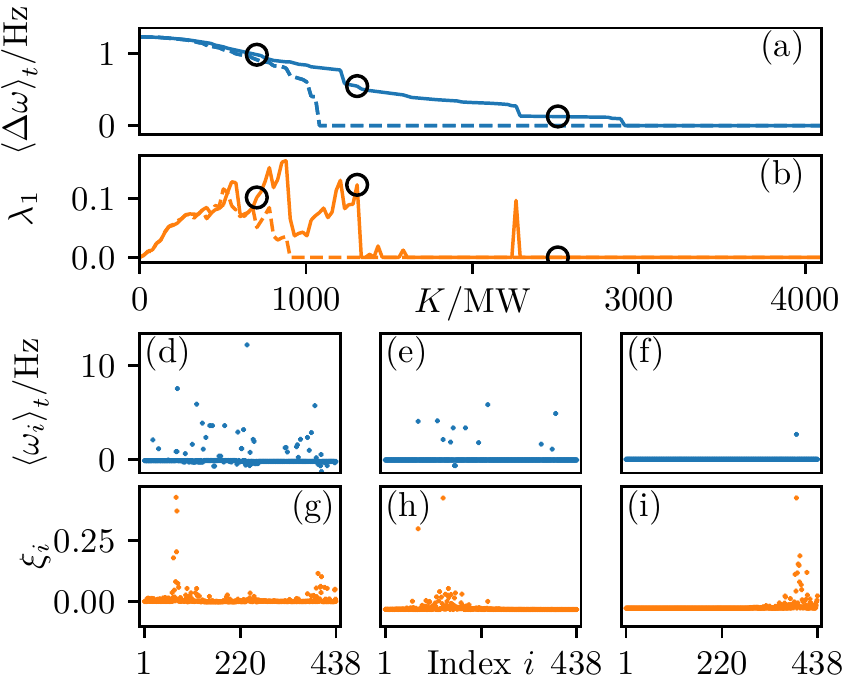}
	\caption{German power grid with real-world distribution $P^\mathrm{R}$ using the extended model: (a) Average frequency deviation $\avg{\Delta \omega}_t$ and (b) largest Lyapunov exponent 
		$\lmax$ versus coupling strength $K$. The solid lines (dashed lines) correspond to the adiabatic upsweep (downsweep) of $K$. \mbox{(c) - (e)} Average frequencies $\avg{\omega_i}_t$ 
		and \mbox{(f) - (h)} Lyapunov vector components $\xi_i$ versus node index $i$ for the $K$ values marked by black circles in the top panels from left to right. Parameters: \mbox{$m_v=1$}, 
		\mbox{$E_{f,i}=1$}, \mbox{$X_i=1$}. Lyapunov exponents and vectors calculated for a duration of $8\cdot 10^3$~s. Lyapunov exponents are expressed in units of $\Delta t^{-1}=5 s^{-1}$.
		Other parameters as in Fig. \ref{fig:adynamicswoutctrlgerrealworld}.}
	\label{fig:avdynamicswoutctrlgerv}
\end{figure}
As for the previously investigated setups, the system undergoes a hysteretic transition to synchronization (see Fig. \ref{fig:avdynamicswoutctrlgerv}a). It shows an asynchronous state
for low coupling values $K$, and partially synchronized states for intermediate $K$ values (panels d, e). In particular the number of whirling nodes diminishes with increasing $K$ and it
is possible to identify a state, in proximity of the synchronization transition, where almost all nodes are synchronized, while few of them are solitary nodes still oscillating with average frequency 
different from zero (panel e). Similarly to the previous setups, the Lyapunov vector is (mostly) localized around solitary nodes (see Fig. \ref{fig:avdynamicswoutctrlgerv}, panels (c)-(h) corresponding to
different stages of the adiabatic upsweep), thus indicating that solitary nodes are leading the synchronization transition even when considering voltage dynamics.
Finally, the system is chaotic for a larger $K$ interval (see panel b) as compared to the original Kuramoto model with inertia.

\begin{figure}[h]
	\centering
	\includegraphics[width=1\linewidth]{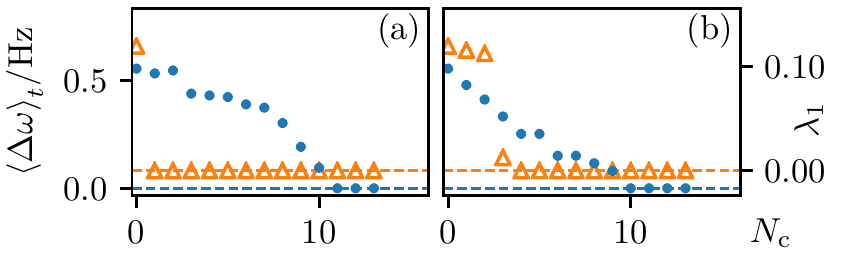}
	\caption{Efficiency of time-delayed feedback control for the German power grid with real-world distribution $P^\mathrm{R}$ using the extended model: time averaged frequency
		deviation $\deva$ (blue dots) and maximum Lyapunov exponent $\lmax$ (orange triangles) vs. number of controlled nodes $\ncontrolled$ following different control strategies: 
		(a) solitary nodes sorted in descending order of $\xi_i$; (b) solitary nodes sorted in descending order of $\abs{\avg{\omega_i}_t}$. The dashed lines mark $\deva=0$, $\lmax=0$. 
		Control acts for a duration of 40 seconds and is then turned off; delay time \mbox{$\tau=4$~s}, feedback gain $g=1.3$, $K\approx 1307$~MW (middle point of Fig. 
		\ref{fig:avdynamicswoutctrlgerv}a). Other parameters as in Fig. \ref{fig:avdynamicswoutctrlgerv}.}
	\label{fig:estrategycomparisongerv}
\end{figure}

Strategies (i) and (ii) to synchronize and stabilize the system  are applied to the partially synchronized state at $K\approx 1307$~MW (see Figure \ref{fig:avdynamicswoutctrlgerv} panels e, h), 
where 13 solitary nodes are present: a comparison of the strategies is shown in Fig. \ref{fig:estrategycomparisongerv}.
The first strategy requires to control one node in order to stabilize the system and 11 to synchronize, whereas the second strategy performs worse when stabilizing the system (4 nodes required) 
but performs better when synchronizing (10 nodes). However both control schemes require not all solitary nodes to be controlled in order to achieve synchronization and stability. 
All in all our approach is not only applicable to the example systems presented in Sec. 3, but works for different models. In Appendix B, the generality of the approach will be further explained
considering different topologies and different operating points. Even though it is not possible to provide an analytical proof of the efficiency and generality of our control approach, our results indicate how powerful and robust time-delayed feedback control is, and that it can be applied to a diversity of topologies and power grid models. The hysteretic nature of the transition to synchronization, the bistability of the system, and the emergence of solitary states driving the dynamics, are fundamental ingredients for enhancing the stability of power grids, which have not been recognized until now.
	
	\section{Conclusions}

	In conclusion, we have proposed a time-delayed feedback control scheme to restore frequency synchronization and stability of the power grid after perturbations. 
		To this purpose we have firstly studied the Kuramoto model with inertia in the presence of two different bimodal distributions of generator and load power (an artificial distribution, and one adapted from the real German high-voltage transmission grid), which both lead to a fully frequency synchronized, stable network for large transmission capacities $K$.
        We have focussed on the operating regime of intermediate $K$ characterized by a number of solitary nodes whose mean frequency deviates from that of all other nodes. 

	We have shown that stability and synchronization can be enhanced by time-delayed feedback control in this $K$ regime by applying delayed feedback to a small subset of nodes: frequency synchronization and stability can be restored in a short time and persist even if control is turned off. 
	Different control strategies were tested. For the shown setup the best strategy is to control the most unstable solitary nodes, characterized by the largest Lyapunov vector components.
	However, both strategies (i) and (ii) are efficient, being based on the solitary nodes that turn out to be fundamental in regulating the dynamics of the system. Solitary nodes exhibit independent dynamics, giving rise to low-dimensional chaos that turns into high-dimensional quasi-periodic motion when the most unstable node is controlled, until synchronization is achieved. 
	Therefore, due to their independence, the set of controlled nodes cannot be much smaller than the number of solitary nodes.
	
	The proposed fast-acting control method might offer an interesting approach to cure disturbances in real-world power grids, due to its general applicability and validity, as shown in Sec. IV,
	where we have applied our control strategy to a more sophisticated model including the voltage dynamics \cite{SCH14m} and, more in general, as shown in Appendix B, where we have extended our analysis to a different network (i.e., the Italian grid) and to different operating points, keeping the German grid topology.

\acknowledgments
We acknowledge A. Torcini and S. Lepri for valuable discussions. 
Funded by the Deutsche Forschungsgemeinschaft (DFG, German Research Foundation) - Projektnummer 163436311 - SFB 910.

\section*{Appendix A: Parameter choice}



As already detailed in Sec. II A, the Kuramoto model with inertia describes the phase and frequency dynamics of $N$ coupled synchronous machines, i.e., generators or consumers within the power grid, 
where mechanical and electrical phase and frequency are assumed to be identical. The $N$ dynamic equations
describing the time evolution of the phase $\ti(t)$ and frequency $\tid(t)=\frac{d\ti}{dt}$ of node $i=1,...,N$ are given by Eq. \ref{eq:kura}. 
In particular $\alpha$ represents the dissipation parameter and takes typical values of 0.1-1 s$^{-1}$ \cite{MEN14,MAC08a}. However, in a realistic power grid there are additional sources of dissipation, especially Ohmic losses, and losses caused by damper windings \cite{MAC08a}, which are not taken into account directly in the coupled oscillator model. Therefore, for this parameter we have chosen slightly higher values: \mbox{$\alpha=5/6$~s$^{-1}$} when a bimodal Gaussian distribution is considered and \mbox{$\alpha=2$~s$^{-1}$} when the real-world distribution is taken into account to describe 
the distribution of the net power $P_i$. Different dissipation values are necessary for the different distributions in order to obtain comparable setups , i.e., 
unstable, partially synchronized states at comparable coupling strengths, \mbox{$K=819$~MW} for the bimodal Gaussian distribution and \mbox{$K=729$~MW} for the real-world one.

For both net power distributions, the coupling strength $K$, which represents the maximum power transmission capacity of transmission lines, was set homogeneously throughout the grid. 
A more realistic approach would have been to use a coupling matrix $K_{ij}$, containing not only the topology, but also individual transmission capacities to schematize different transmission line lengths. 
However the goal of the present paper is to gain insight into the principal behavior of large power grids depending on the network
topology, and their capability to synchronize by controlling a minimal set of nodes and, for a proof of principle of our control approach, the choice of identical transmission lines suffices. 
The choice of using simplified homogeneous transmission line capacities (coupling constants) turned out to be a good compromise 
when using heterogeneous power distributions, whose realistic values were available in the open data source as opposed to the power distribution data. 

Eq. (\ref{eq:kura}) can be simplified by rescaling the parameters \mbox{$m:=\frac{1}{\alpha}$}, \mbox{$\Omega_i:=\frac{P_i}{I_i\omgg \alpha}$}, \mbox{$k:=\frac{K}{I_i\omgg \alpha}$},
thus giving
\begin{align}
m\tidd+\tid=\Omega_i + k\sum_{j=1}^{N}\amat\sin(\tj-\ti).\label{eq:kurand}
\end{align} 
In comparison with Eq.(\ref{eq:kura}), the inertial mass $m$ now represents the inverse of the dissipation $\alpha$ in the grid, and the coupling constant $k$ now represents the maximum power which can be transmitted between two connected nodes. Moreover each node $i$, when uncoupled, oscillates with an angular frequency $\Omega_i$, referred to as \textit{natural frequency} or \textit{inherent frequency}. Therefore the distribution of natural frequencies and the distribution of net power $P_i$ are equivalent, up to a constant factor.

Finally, adiabatic simulations (upsweep of $k$) are performed to measure the level of synchronization in the network starting from the asynchronous state towards the partially synchronized state.
In particular the rescaled coupling strength $k$ is increased from $k=0$ to $k=60$ in steps of $\Delta k=0.4$ (from $k=0$ to $k=60$ in steps of $\Delta k=0.2$) for the bimodal Gaussian distribution (real-world distribution, respectively).   
Specifically, for the bimodal Gaussian distribution with \mbox{$\alpha=5/6$~s$^{-1}$}, \mbox{$I_i=40\times10^3 \text{kg}\text{ m}^2$}, \mbox{$\omgg=2\pi \cdot 50$~Hz} and $\Delta k=0.4/\Delta t$ one obtains $\Delta K = \Delta k I_i  \omgg  \alpha \approx 21$~MW, if a time unit $\Delta t=0.2$ s is considered.

\section*{Appendix B: Generality of the results}
In order to show that the efficiency of our proposed control strategies is not restricted to the setups shown in the main text, we will present additional results: (a) 
keeping the setups shown in the main text, but analyzing different operating points and different configurations by considering different coupling strengths; 
(b) taking into consideration a different topology.

\subsection*{Different operating points}
In this section we present the results for a different operating point, thus giving rise to a different configuration of solitary nodes.
In particular, keeping the same setups presented in the main text, we show a comparison between the strategies (i) and (ii) obtained when the system is evaluated at
different coupling strengths, thus investigating different working points with respect to the results shown in the main text. 
For the bimodal Gaussian distribution $P^G$, we investigate the state at $K\approx 565$~MW, which is a partially synchronized state found during the upsweep protocol, characterized 
by 19 solitary nodes. This configuration is unstable, with $\lambda_1=0.0144\pm 0.0005$. 
Regarding the real-world distribution $P^R$, the different working point that we have investigated is characterized by $K\approx 578$~MW, 19 solitary nodes and $\lambda_1=0.116\pm 0.005$. 

\begin{figure}[h]
	\centering
	\includegraphics[width=1\linewidth]{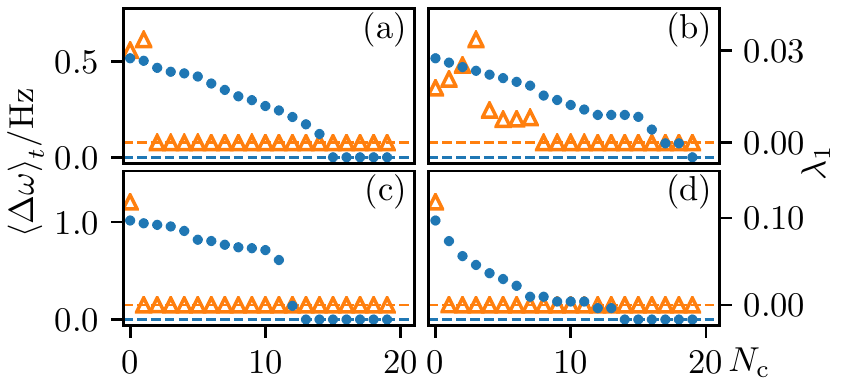}
	\caption{Efficiency of time-delayed feedback control: time averaged frequency deviation $\deva$ (blue dots) and maximum Lyapunov exponent $\lmax$ (orange triangles) vs. 
		number of controlled nodes $\ncontrolled$ following different control strategies: (a), (c) solitary nodes sorted in descending order of $\xi_i$; (b), (d) 
		solitary nodes sorted in descending order of $\abs{\avg{\omega_i}_t}$. Panels (a)-(b) correspond to the distribution $\Pgauss$, (c)-(d) to $\Preal$. 
		The dashed lines mark $\deva=0$, $\lmax=0$. Control acts for a duration of 40 seconds and is then turned off; delay time \mbox{$\tau=4$~s}, feedback gain $g=1.5$ $(g=1)$ for 
		$P^G (P^R)$. 
		Other parameters as in Fig. \ref{fig:adynamicswoutctrlgergauss} for the top panels (Fig. \ref{fig:adynamicswoutctrlgerrealworld} for the bottom panels).} 
	\label{fig:estrategycomparisongerdiffK}
\end{figure}
The outcome of the control schemes is shown in Fig. \ref{fig:estrategycomparisongerdiffK}. For the $P^G$ distribution strategy (i) requires the control of 2 solitary nodes to stabilize
the system and 15 to synchronize, while strategy (ii) requires the control of 8 nodes to stabilize and 19 to synchronize the system.
For the realworld distribution both strategies require one controlled node to stabilize. Synchronization is reached with 13 and 14 nodes using strategy (i) and (ii) respectively.

\subsection*{Italian grid}
\begin{figure}
	\centering
	\includegraphics[width=0.9\linewidth]{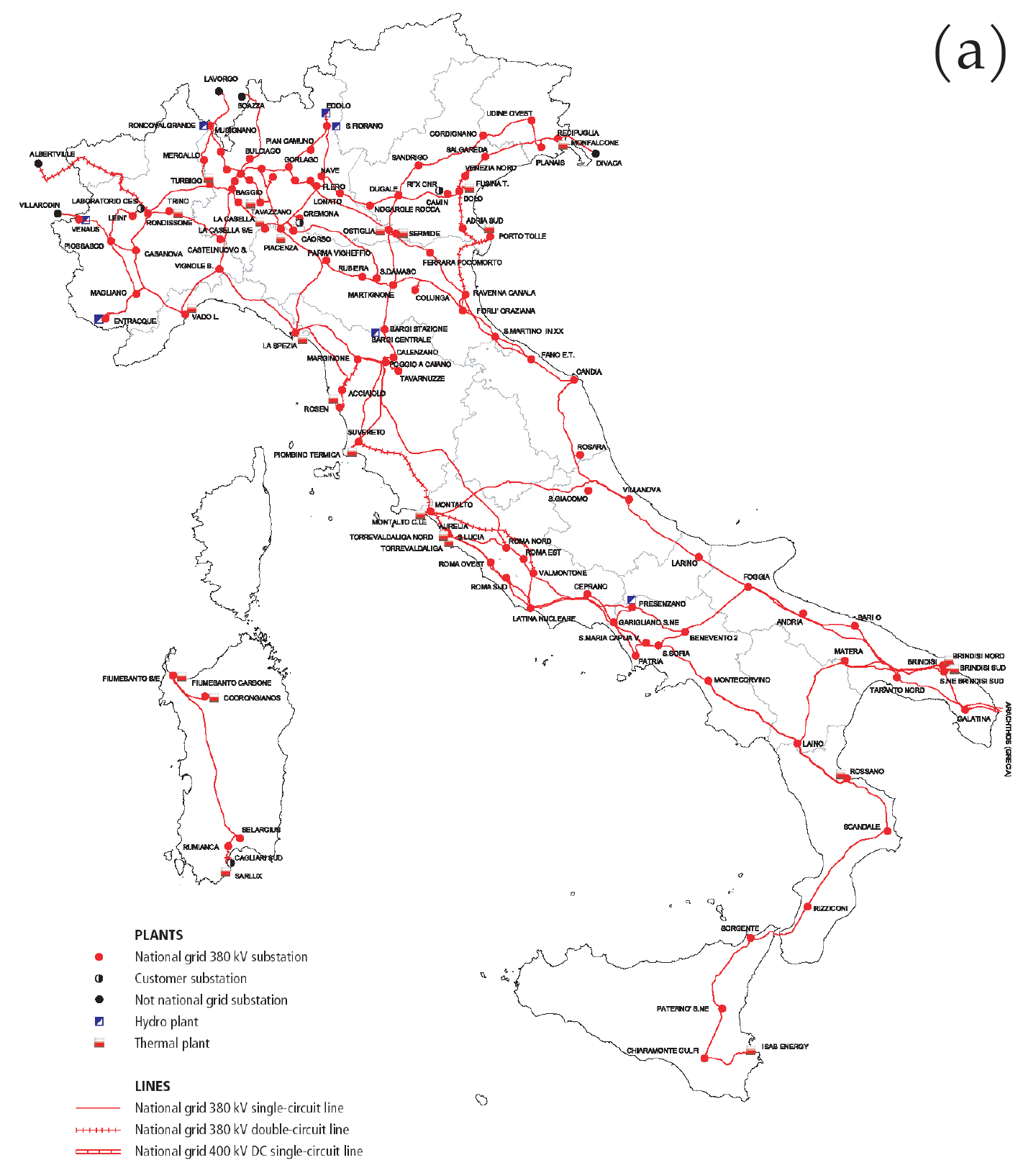}
	\includegraphics[width=1\linewidth]{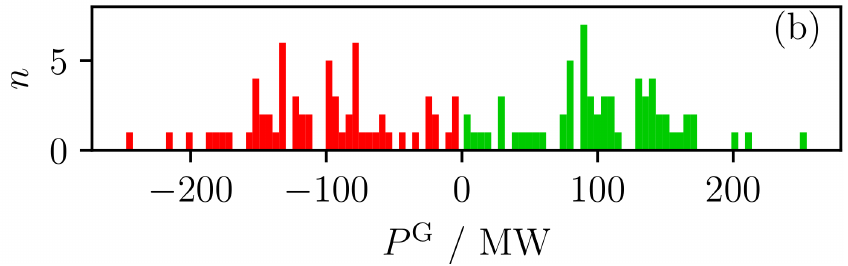}
	\caption{(a) Map of the Italian ultra-high voltage power grid, consisting of 127 nodes connected by 171 transmission lines (red lines) \cite{ita}. 
		(b) Histogram shows a realization of an artificial bimodal Gaussian distribution of net power with $N=127$; $P_0=105$~MW, $\sigma=P_0/2$.}
	\label{fig:0distriita}
\end{figure}

In this section we apply our control strategy to a different grid topology. The dynamics of the single node is still described by Eq.~(\ref{eq:kura}), but we now consider the Italian high-voltage (380 kV) power grid (Sardinia excluded), which is composed of N = 127 nodes, divided into 34 generators (hydroelectric and thermal power plants) and 93 consumers,
connected by 171 transmission lines \cite{ita}. This network is characterized by a quite low average connectivity $\left\langle  N_c\right\rangle = 2.865$, due to the geographical distributions of the 
nodes along Italy (see Fig. \ref{fig:0distriita}a). Since we have no access to a distribution of generator powers and nodal power consumption, we restrict the application of our 
method to the artificial distribution, using a bimodal Gaussian distribution (shown in Fig. \ref{fig:0distriita}b) with the same probability density function as the one used for the German grid (see Eq. \ref{p_gaussian} of the main text).

\begin{figure}
	\centering
	\includegraphics[width=1\linewidth]{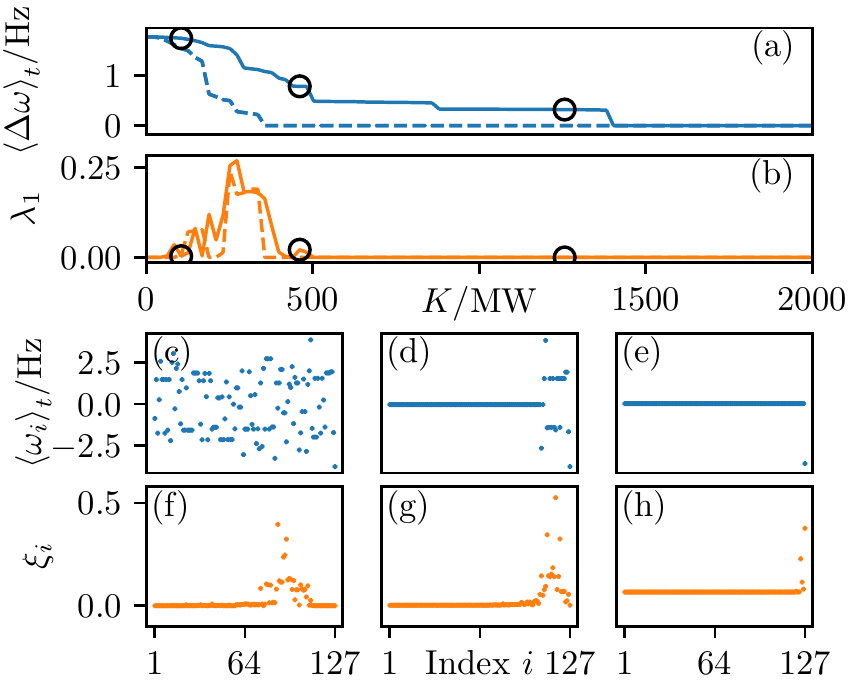}
	\caption{Italian power grid with Gaussian distribution $P^\mathrm{G}$: (a) Average frequency deviation $\avg{\Delta \omega}_t$ and (b) largest Lyapunov exponent $\lmax$ versus coupling 
		strength $K$. The solid lines (dashed lines) correspond to the adiabatic upsweep (downsweep) of $K$. \mbox{(c) - (e)} Average frequencies $\avg{\omega_i}_t$ and \mbox{(f) - (h)} 
		Lyapunov vector components $\xi_i$ versus node index $i$ for $K$ values marked by the black circles in the top panels from left to right. Lyapunov exponents and vectors calculated for a 
		duration of $2\cdot 10^4$. Other parameters as in Fig. \ref{fig:adynamicswoutctrlgergauss}.}
	\label{fig:adynamicswoutctrlitaly}
\end{figure}

Like for the German grid, the synchronization transition is hysteretic (see Fig. \ref{fig:adynamicswoutctrlitaly}a), but the formation of frequency clusters at different stages of the 
upsweep protocol is more pronounced since the local architecture favours a splitting based on the proximity of the oscillators.
At $K\approx 461$~MW (middle black point of Fig. \ref{fig:adynamicswoutctrlitaly}a) the system is partially synchronized and unstable ($\lmax>0$): it represents a big cluster of locked
oscillators with zero average frequency and 20 unsynchronized whirling oscillators (see panel d). Besides the main frequency-synchronized cluster, two other clusters can be found: one with 
positive and one with negative average frequency, consisting of eight and five nodes, respectively. The remaining seven nodes are solitary. As before, we will take this state as 
an example to be controlled using our proposed strategies.
For smaller coupling the system is unstable, but completely asynchronous (see panels b, c), while for larger coupling the system is (almost) completely synchronized (see panel e): one solitary
state corresponding to the last node in Sicily hardly synchronizes due to the peripheric position in the network.

\begin{figure}
	\centering
	\includegraphics[width=1\linewidth]{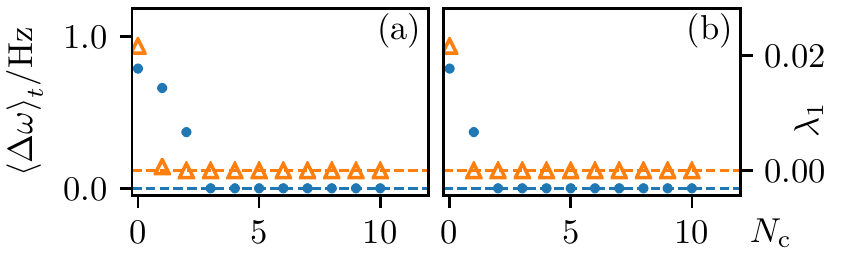}
	\caption{Efficiency of time-delayed feedback control for the Italian grid: time averaged frequency deviation $\deva$ (blue dots) and maximum Lyapunov exponent $\lmax$ (orange triangles) vs. 
		number of controlled nodes $\ncontrolled$ following different control strategies: (a) solitary nodes sorted in descending order of $\xi_i$. (b) solitary nodes sorted in descending 
		order of $\abs{\avg{\omega_i}_t}$. The dashed lines mark $\deva=0$, $\lmax=0$. Control acts for a duration of 40 seconds and is then turned off; delay time \mbox{$\tau=4$~s}, 
		feedback gain $g=1$, $K\approx 461$~MW (middle point of Fig. \ref{fig:adynamicswoutctrlitaly}). Other parameters as in Fig. \ref{fig:adynamicswoutctrlitaly}.}
	\label{fig:estrategycomparisonitaly}
\end{figure}

In Fig. \ref{fig:estrategycomparisonitaly} a comparison of strategies (i) and (ii) is presented.
First of all, as for the German grid, the delayed feedback control is able to synchronize and stabilize the grid when enough nodes are controlled. Strategy (i), which controls preferably the most
unstable nodes, sorted according to their Lyapunov vector component $\xi_i$, needs two nodes to stabilize and three controlled nodes to synchronize the system (see panel a). 
On the other hand, by employing strategy (ii), which orders the controlled nodes with respect to their frequency deviation $\abs{\avg{\omega_i}_t}$, the control of one node is required to stabilize, 
and two controlled nodes to synchronize the system (panel b). In both cases a remarkably small fraction of the 20 whirling nodes has to be controlled to gain the suitable conditions for 
operating power grids, thus highlighting the role played by solitary nodes in driving the network dynamics.


\begin{thebibliography}{37}
\bibitem {pert}	It is important to notice that we do not explicitly perturb the system, but we consider a partially synchronized state naturally coexisting
	with the synchronized one due to the hysteretic nature of the synchronization transition.
	\BibitemOpen
\bibitem{parameter} The different values for the system with bimodal Gaussian distribution $P^G (\alpha = 5/6\enskip s^{-1})$ and real-world distribution $P^R (\alpha = 2\enskip s^{-1} )$ are chosen in order to find unstable, partially synchronized states at comparable coupling strengths, $K = 819$ MW and $K = 729$ MW, respectively.	
	\BibitemOpen
\end{thebibliography}
\end{document}